\documentclass[12pt]{article}
\usepackage{amssymb} 
\usepackage{amscd} 

\newcommand {\newsection}{\setcounter{equation}{0}\section}

\newcommand{\Tr}{\mathop{\rm Tr}\nolimits}
\newcommand{\Id}{\mathop{\rm Id}\nolimits}

\newcommand{\leftpartial}{\mathop{\!\stackrel{\leftarrow}{\partial}}\nolimits}
\newcommand{\rightpartial}{\mathop{\!\stackrel{\rightarrow}{\partial}}\nolimits}

\newcommand{\ket}[1]{|#1\rangle}

\let\eps = \epsilon

\def \l {\lambda}
\def \be {\begin{equation}}
\def \ee {\end{equation}}
\def \ba {\begin{array}}
\def \ea {\end{array}}
\def \bea {\begin{eqnarray}}
\def \eea {\end{eqnarray}}

\def \rr {{\mathbb R}}
\def \hh {{\mathbb H}}

\def \aa {{\cal A}}
\def \nc {noncommutative }
\def \ncg {noncommutative geometry }
\def \da {\dagger}

\begin{document}

\baselineskip 0.65 cm
\begin{flushright}
SISSA 64/00/EP/FM\\
hep-th/0006091
\end{flushright}
\begin{center}

{\Large{\bf Noncommutative $SO(n)$ and $Sp(n)$ Gauge Theories}}
\vskip .5cm

{\large L. Bonora$^{1,3}$, M. Schnabl$^{1,3}$, M.M. Sheikh--Jabbari$^2$, 
A. Tomasiello$^1$}

\vskip .5cm
{bonora@sissa.it, schnabl@sissa.it,jabbari@ictp.trieste.it,tomasiel@sissa.it}
\vskip .5cm

 {\it $^1$ Scuola Internazionale Superiore di Studi Avanzati,
 
Trieste, Via Beirut 2, 34014 Trieste, Italy}

 {\it $^2$ The Abdus Salam International Centre for Theoretical 
Physics\\
Strada Costiera 11, 34014 Trieste, Italy}\\

 {\it $^3$ INFN, Sezione di Trieste}
\end{center}

\vskip 2cm 
\begin{abstract}
We study the generalization of \nc gauge theories to the case of
orthogonal and symplectic groups. We find out that this is possible,
since we are allowed to define orthogonal and symplectic subgroups 
of \nc unitary gauge 
transformations  even though the gauge potentials and gauge 
transformations are not valued in the orthogonal and symplectic 
subalgebras of the Lie algebra of antihermitean matrices. Our 
construction relies on an antiautomorphism of the basic \nc 
algebra of functions which generalizes the charge conjugation 
operator of ordinary field theory. We show that the corresponding 
\nc picture from low energy string theory is obtained via orientifold 
projection in the presence of a non--trivial NSNS B--field. 
\end{abstract}
\newpage   
\newsection{Introduction}
It is well-known that a constant NSNS two form B-field background  can
be gauged away in the perturbative type II string theories. The addition of
D-branes drastically modifies this conclusion.  
The novel point is that the components of a constant 
B-field background which are parallel to a $Dp$-brane can not be gauged away
anymore \cite{{Bound},{AAS},{Chu}}. Studying the open strings ending 
on D-branes with B-field turned on, it has been shown that the worldvolume 
of these branes become noncommutative \cite{{AAS},{Chu},{Dir}}.
In addition, by computing open string states 
scattering amplitudes and extracting the massless poles contributions, one
can show that the low energy effective theory describing the system
is the noncommutative $U(1)$ ($NCU(1)$) theory \cite{{HD},{Sh}}.  
  
In the zero B-field background we know that the low energy effective theory 
of $n$ coincident $Dp$-branes is a $p+1$ dimensional supersymmetric (with 16 
supercharges) $U(n)$ theory. 
The $U(1)$ part of this $U(n)$ basically represents the
interactions of D-brane open strings with the bulk closed strings (supergravity 
fields). This $U(1)$ part, which is usually called the center of mass $U(1)$, 
decouples from the open strings dynamics and hence effectively we find an
$SU(n)$ theory. However, for $n$ coincident
D-branes in a constant B-field background, the above argument is modified by 
the fact that the brane worldvolume is a noncommutative Moyal plane. In this 
case we deal with a noncommutative version of $U(n)$ theory, namely, $NCU(n)$. 
This theory is obtained by replacing the usual products of fields by the 
star (Moyal) product \cite{SW}.
However, in this case separating the center of mass (noncommutative) $U(1)$ is 
impossible.  Intuitively, this corresponds to the fact that when
we turn on a B-field, the open string left and right movers (holomorphic and
anti-holomorphic parts) contribute {\it unequally}, but in the closed strings 
sector 
(bulk fields) left and right movers always appear on an equal footing.  So,
in view of the open--closed strings interactions, we cannot decouple $U(1)$ 
modes. They contribute a part, which, as expected, 
depends on the background B-field. Apart from this string theoretic
reasoning, one can also understand this point through a gauge theory argument. 
Suppose we start with the usual $SU(n)$ gauge theory and make it noncommutative by
replacing the products with star products. Then one can show that this theory 
will not be consistent: the \nc gauge transformations will not close to form 
a group. In other
words, by performing a \nc $SU(n)$ gauge transformation we create 
extra
terms which 
can consistently appear only in 
a \nc $U(n)$ theory \cite{Adi}. 

As one can see from the above discussion, trying to define a \nc gauge theory
corresponding to subgroup of $U(n)$ and a string/brane theory configuration 
that corresponds to it, does not look very promising, at first sight. 
However we will show below that, to a certain extent, it is possible to find
consistent \nc extension for gauge theories corresponding to  
certain subgroups of $U(n)$. The main observation is that it is
possible to define gauge transformations that close to form a subgroup
of the group of $NCU(n)$ gauge transformations even though the 
corresponding gauge potentials and gauge transformations are not valued
in a classical Lie subalgebra of the unitary Lie algebra $u(n)$.
We handle this problem from two different points of view, one 
relying on purely gauge field theory considerations and the other
on string 
theory. From the gauge theory point of view, we show that it is 
possible to impose constraints on the gauge potentials and the gauge 
transformation so that when the deformation parameter vanishes we 
recover the ordinary orthogonal and symplectic gauge theories. 
An essential role in defining the constraints is played by an 
operator which is a generalization of the charge conjugation 
operators for gauge theories.
 
{}From a string theory point of view, the analogous operator is that 
which in the ordinary cases (without background B-field) is the 
worldsheet parity (possibly plus spatial parity) which is responsible 
for orientifold projections. So, a natural framework from a string 
theory point of view seems to be an orientifold in the presence of D-branes 
with a B-field switched on.

The paper is organized as follows. In section 2, we briefly review the \nc
gauge theories and show that, in general, the concept of \nc gauge group should 
be modified compared to commutative theories. We will argue that one should 
abandon the naive picture of connections taking their values in a Lie algebra 
of the gauge group. The generalization we need is based on an antiautomorphism,
$r$, in the corresponding C$^*$-algebra of functions. We show that this 
$r$ map together with the matrix transposition is a realization of the
charge conjugation operator we need to extract $SO$ and $Sp$ parts out of 
$U(n)$.

In section 3, we present our string theoretic arguments. First we discuss the 
issue
of orientifold planes with a particular step function-like B-field, which 
passing through the O--plane changes the sign and is zero on the plane. 
Putting $Dp$-branes in this background parallel to $Op$-planes, we study the 
supersymmetries preserved
by this system and we show that it is stable. Then, we proceed to find the low 
energy effective action for open strings attached to such $Dp$-branes. We show 
that actually it coincides with the results of our field theory arguments. 
Last section is devoted to remarks and further discussion.

\newsection{$SO(n)$ and $Sp(n)$: \nc realization}
\label{math}

\subsection{Preliminary discussion}

Before we describe our proposal, it is useful to illustrate what are the 
problems connected with the realization of noncommutative orthogonal and
symplectic groups. This subsection is pedagogical in nature and the reader
may wish to skip it. The definition
and treatment of noncommutative orthogonal and symplectic groups start 
in the next subsection.

First, let us discuss what we mean by gauge group in \nc geometry. 
Since we have in mind the case of $\rr^d$, let us use
the Moyal approach (even though some of the remarks which follow are
independent of this assumption) and define our starting complex algebra 
$\aa_{\theta} $ as the vector space
$C^\infty(\rr^d)$ endowed with product
\be
f*g(x) \equiv f(x) e^{\frac i2 \theta^{\mu\nu}\leftpartial_\mu 
\rightpartial_\nu}g(x).
\label{Moyal}
\ee
Let now $A$ be the connection (the rank is not necessarily 1; the indices are
understood), which transforms under gauge transformation as $A_U = 
U^{-1} *(d + A*) U$, where $U^{-1}*U=1$. The hermiticity condition $A^{\dagger}= 
- A$ is preserved if $U^\da= U^{-1}$; these $U$'s are called  
unitary automorphisms (of the module on which the gauge theory is
constructed), but {\sl they are not} functions on $\rr^d$ valued in $U(n)$. 
It is true instead that the connection $A$ and 
the infinitesimal gauge transformations $\lambda$ are $u(n)$--valued functions
on $\rr^d$. Indeed, they are both antihermitean:
being $\delta A = d\lambda + A * \lambda
 - \lambda * A$, the antihermiticity condition 
on $A$ is preserved if also $\lambda^\da =  - \lambda$ holds, thanks to
the relation 
\be
(f * g)^{\dagger} 
= g^{\dagger} * f^{\dagger},
\label{staralg}
\ee 
which is a generalization of the $*$-algebra property 
to matrix--valued functions. 
For the same reason $F$, the field strength,
\be
F_{\mu\nu}=\partial_{[\mu}A_{\nu]}+A_{\mu}*A_{\nu}-A_{\nu}*A_{\mu},
\ee
is antihermitean too. 

Now, when dealing with orthogonal and symplectic groups, something worse
happens. The connection is no longer a function from $\rr^d$ to 
${\rm Lie}(G)$; as we briefly discuss below, this is impossible to
accomplish. But, what we said above should convince the reader that 
in \ncg this is not such
a dramatic loss: finite gauge transformations are just unitary automorphisms
of the module, and their infinitesimal counterparts are $U(n)$-valued for the
simple reason that there is no product in the antihermiticity condition. 
Our theory is a \nc gauge theory which reduces in the commutative limit to the
desired one, and, what is most important, seems to have a physical origin in
string theory.

Let us now discuss in some detail why one has to abandon the 
${\rm Lie}(G)$-valuedness of the gauge potential. 
{}From the description above, one would hope to generalize $SO(n)$
and $Sp(n)$ gauge groups in a simple way: just replace the complex (because of the
$i$ in (\ref{Moyal})) algebra we
started with by a real (resp. quaternionic) one.
Let us illustrate such an approach with the example $Sp(1) = SU(2)$. 
Could we find a deformation 
of the algebra $\aa \equiv C^\infty(\rr^d, \hh)$ 
of functions 
from $\rr^d$ to $\hh$ (quaternions), we would be done. 
Elements of $\aa$ can be expressed as $f= f_i \tau_i$. The idea of
tensoring the usual 
Moyal product (\ref{Moyal}) with the matrix one (a definition
like $(f_i \tau_i) * (g_j \tau_j) \equiv (f_i * g_j) \ \tau_i \tau_j\,$) is
clearly too naive:
this would not be a product in $\aa$, since it does not close. If we
start, indeed, 
from real $f_i$ and $g_j$, we get a complex $f_i * g_j$ (recall  
the $i$ in 
(\ref{Moyal}), which cannot be dropped because it is needed for the
property (\ref{staralg}) and hence for gauge symmetry). 

Alternatively one can exhibit true products. The easiest try
would be to consider the $i$ in (\ref{Moyal}) as one of the imaginary
quaternions $i$, $j$, $k$. In this way the property (\ref{staralg})
remains valid, but what is lost is associativity. 
Although this initial try is wrong, it illustrates the idea. 
However, we have checked to first order 
that there is no associative deformation compatible with (\ref{staralg}).
Even more, suppose one wants 
to take a pragmatic point of view and accept to live with
non associative products. Then one may look directly for deformed
Moyal brackets which satisfy Jacobi identity. It turns out that at first 
order such brackets do not exist.

In conclusion, if we want to find a realization of noncommutative orthogonal 
and symplectic groups, we seem to be obliged to give up the familiar idea of
Lie($G$)--valued connections.

\subsection{Noncommutative $SO(n)$ and $Sp(n)$}

We are now ready to illustrate our proposal for the generalization of
orthogonal and symplectic group. 

\subsubsection{The $r$ antiautomorphism}

To start with, we will work in a setting in which $\theta$ has to be thought
of as a parameter. Accordingly, we will consider $\aa_\theta$ 
as an algebra of (possibly formal) power series in $\theta$. This algebra has 
an anti--automorphism $r$ defined by
\be
(.)^r: 
f(x,\theta) \mapsto f^r(x,\theta) \equiv f(x, -\theta). \nonumber
\ee
This map reduces to the identity on the generators $x^\mu$ and reverses 
the order in 
the product: $(x^\mu_1 * \ldots * x^\mu_n)^r
= (x^\mu_n)^r * \ldots * (x^\mu_1)^r$. 
 
\subsubsection{Orthogonal and symplectic constraints}

First of all, we consider our groups as subgroups of $U(n)$. In other words 
we keep the usual antihermiticity condition on the $u(n)$--valued connections 
$A$ and gauge transformations $\lambda$. To fix our conventions we will use Greek
letters for space-time indices and $i$ and $j$ for matrix (group) indices. 
Here, for later use, we write down explicitly the hermiticity condition:
 \begin{eqnarray}\label{herm}
A_{ij}^*(x,\theta) &=& - A_{ji}(x,\theta)  \nonumber\\
\lambda_{ij}^*(x,\theta) &=& - \lambda_{ji}(x,\theta).\label{reality}
\end{eqnarray}

Our defining condition for the $NCSO(n)$ connections and gauge transformations 
is to take the gauge connections and transformations satisfying the following
constraints:
\begin{eqnarray}\label{const}
A^r_{ij}(x,\theta) &=& - A_{ji}(x,\theta)  \nonumber\\
\lambda^r_{ij}(x,\theta) &=& - \lambda_{ji}(x,\theta)\label{ortcond}
\end{eqnarray}
 
Let us comment on these constraints. First of all, it is easy to see that they
are preserved by gauge transformations. One can see it componentwise.
Alternatively, rewrite (\ref{ortcond}) in the concise form $A = - (A^t)^r$
and $\lambda = - (\lambda^t)^r$, i.e. $t$ is the matrix transposition. 
Define $(\,(.)^t)^r\equiv (.)^{rt}$; one can show that the $rt$ map 
enjoys the (\ref{staralg}) property, with $rt$ replacing $\dagger$. 
The proof is now formally similar to the usual one for 
$U(n)$: $(\lambda * A - A * \lambda)^{rt} = 
A^{rt} * \lambda^{rt}  - \lambda^{rt}* A^{rt} = 
- (\lambda * A - A * \lambda)$.

The second comment we wish to make is that 
the constraints we introduced are natural if one recalls that in \nc gauge
theories the map $-(\cdot)^{rt}$ is nothing but complex conjugation; our
theory is the charge-conjugation invariant version of the usual one. 
More explicitly, as discussed in \cite{CPT}, indeed the charge conjugation operator
is
\be
A^c=-A^{rt}.
\ee
One can write an explicit solution of (\ref{const}) as:
\be\label{ncso}
{\cal A}_{\mu}=\frac 12(A_{\mu}-A_{\mu}^{rt})=\frac 12(A_{\mu}+A_{\mu}^c).
\ee
This notation may be ambiguous and we hasten to specify that when (\ref{ncso})
is used we understand that $A_\mu$ transforms with gauge parameter
$\Lambda=\frac 12(\lambda- \lambda^{rt})$\footnote{In the ordinary commutative case,
this is the way to `reduce' a unitary connection to an orthogonal one, 
\cite{KN}, Prop.6.4.}. More precisely, our ${\cal A}_{\mu}$ enjoys the noncommutative
gauge transformations generated by $\Lambda$:
\be
{\cal A}_{\mu}\rightarrow {\cal A}'_{\mu}=U_*^{-1}(\Lambda)*{\cal
A}_{\mu}*U_*(\Lambda)-U_*^{-1}(\Lambda)*\partial_{\mu}U_*(\Lambda),
\ee
where
\be\ba{cc}
U_*(\Lambda)\equiv 1+i\Lambda-{1\over2}\Lambda*\Lambda+.... \\
U_*^{-1}(\Lambda)=U_*(-\Lambda)\ , \ \  U_*^{-1}\ *\ U_*=1. 
\ea\ee

As we see it is immediate that our $NCSO(n)$ gauge fields are charge 
conjugation invariant.
 
Thirdly, we anticipated above that under the (\ref{ortcond}), connections and 
gauge 
parameters do not turn out to be $so(n)$--valued. Nevertheless (\ref{ortcond})
introduces restrictions on the matrix functions $A_{ij}$. To see what they are,
let us write (\ref{ortcond}) more explicitly
\begin{eqnarray}
A_{ij}(x,\theta) &=& - A_{ji}(x,-\theta)  \nonumber\\
\lambda_{ij}(x,\theta) &=& - \lambda_{ji}(x,-\theta)\label{ortcond1}
\end{eqnarray}
Inserting a power expansion in $\theta$ for $A$ 
\begin{equation}
A^\mu(x,\theta) =  A_0^\mu(x) +
i\theta_{\nu\rho} A_1^{\mu\nu\rho} (x) + \ldots,\label{powexp}
\end{equation}
we see that (\ref{ortcond}) implies that $A_0, A_2, \ldots$ are
antisymmetric and $A_1, A_3 \ldots$ symmetric. 
The hermiticity condition (\ref{reality}) imposes that all the coefficients 
$A_0,A_1,...$ be real. The same conclusions hold for the power expansion 
of $\lambda$.

Up to now, $A_0, A_1,\ldots$ are unrestricted, except for the just mentioned
constraint. However, if we want to make connection with string theory, $A_1,
A_2, \ldots$ are expected not to introduce new degrees of freedom, but to be
functionally dependent on $A_0$. The simplest proposal is to regard them as
given by 
the Seiberg--Witten map \cite{SW}:
\be
A^\mu (A_0) = A_0^\mu - \frac i4 \theta^{\nu\rho}
\{ A_{0\nu}, \partial_\rho A_0^\mu + F_{0\rho}^{\ \mu}\}+ {\cal O}(\theta^2);
\label{SW}
\ee
(the presence of $i$ is due the fact that Seiberg and Witten use hermitean
connections rather than anti-hermitean ones, as we do).
This is indeed
consistent: the term linear in $\theta$ is symmetric if the constant
part is antisymmetric. In fact, one can also see that the next term is
antisymmetric, and so on; so we have complete accord with (\ref{powexp}).

This is related to the further subtle issue of 
fixing $\theta$ to a particular value. In this case, of course the approach
we have taken so far --  considering $\theta$ as a formal parameter -- 
loses its validity, and the very definition of $r$ is in jeopardy. However,
thanks to the fact that $A_1, A_2, \ldots$ depend on $A_0$, even when one puts
$\theta$ to a particular value, $A$ is not the most general $U(n)$
field; our constraint becomes more involved but is still there. If we invert the
map to obtain $A_0(A)$, the constraint can be formulated simply as 
\be
A_0(A)
= - A_0^t(A).
\ee 
So we could say that our 
theory is the image of the Seiberg--Witten map restricted to the $SO(n)$ case.
\vskip 0.2cm
It is now easy to introduce similar definitions for noncommutative $Sp(n)$. 
One imposes in this case 
the condition $J A^r = - A^t \, J$, where $J = \eps \otimes \Id_n$, where
$\eps= i \sigma_2$. This constraint is
preserved by gauge transformations that satisfy the same condition.

One could think that the group $SU(n)$ could be tackled in a similar
way: by defining a constraint like $\Tr (A + A^{rt})=\Tr (A + A^r)=0$. 
However, this would not
be gauge invariant.
So, even by using the $r$ map, 
it is not possible to define a $NCSU(n)$ gauge theory.  

\subsubsection{Field theory}

To define a Yang--Mills $NCSO(n)$ theory, 
let $A= A(x, \theta)$ satisfy the constraint (\ref{ortcond}).
The action is the usual one 
\be
S = - \frac14 \int d^d x  { F}_{ij}^{\mu\nu} { F}_{ji\mu\nu}\label{action},
\ee
where  $F$ is defined as 
\be
{ F}_{\mu\nu}=\partial_{[\mu}{ A}_{\nu]}+{ A}_{\mu}*{ A}_{\nu}-
{ A}_{\nu}*{ A}_{\mu}.
\ee
The action (\ref{action}) is naturally gauge invariant under $NCSO(n)$
and positive. It reduces to the usual one for $SO(n)$ in the $\theta =0$ case. 

It is rather straightforward to introduce matter fields in this context in a 
coherent way. For example, suppose we want to introduce fermions in the
adjoint representation. Let us consider a generalization of the Seiberg-Witten map
to such fields. 

Let $\psi_0$ be an ordinary spinor in the adjoint 
representation, which therefore under an ordinary gauge transformations
transforms as follows 
\be
\delta_{\l_0} \psi_0 = [\psi_0,\l_0]\label{cadtr}\ .
\ee
Then it is reasonable to  postulate the following \nc gauge transformation for the
corresponding \nc field:
\be
\delta_{ \l} \psi = \psi*\l-\l *\psi \label{ncadtr}\ , 
\ee
where $ \l = \l_0 + \l'(\l_0,A_0)$, $A = A_0 + A'(A_0)$ and
$ \psi = \psi_0+ \psi'(\psi_0,A_0)$; the primed fields are first order
in $\theta$. We want to find a function $\psi(\psi_0,A_0)$ which transform as
(\ref{ncadtr}) when the corresponding $\psi_0$ transform as (\ref{cadtr}). 
This amounts to satisfying 
the equation
\be
 \psi(\psi_0,A_0) + \delta_{ \l} \psi (\psi_0,A_0) =
 \psi (\psi_0 +\delta_{\l_0} \psi_0 , A_0 +\delta_{\l_0} A_0)\ .
\ee
The solution to first order in $\theta$ is
\be
\psi(\psi_0,A_0) = \psi_0 -\frac i2 \theta^{\mu\nu} 
\left(\{A_{0\mu},\partial_{\nu}
\psi_0\} +\frac 12 \{[\psi_0,A_{0\mu}] ,A_{0\nu}\}\right)+
{\cal O}(\theta^2)\ .
\ee

It is easy to see that our \nc orthogonal constraint
\be
\psi^{rt}= -\psi\label{ortpsi}
\ee
is consistent with this map. Therefore such spinors form a representation of
$NCSO(n)$. 
In a similar way one can introduce also the fundamental representation. The
action terms containing these matter fields are the usual one with
ordinary product replaced by the \nc one, and will not be written down here.

\newsection{Orientifolds and $B$ field.}

\label{phys}
We want now to derive the gauge theory we
described above from a brane configuration in string theory in the limit
$\alpha'\to 0$.

In the commutative case, gauge theories with orthogonal or symplectic groups
are realized as low energy effective actions of branes on orientifold planes
in type I theories. 
Since noncommutativity is achieved by a non zero $B$ field, and this vanishes
on the orientifold plane, one may deem our search hopeless. As an example 
let us consider type IB theory, in which case the gauge groups of the 
branes are the ones we are looking for. However a standard argument tells us 
that the $B$ field is absent, except for a quantized background \cite{BS}; 
for a comment on the quantized $B$ and noncommutativity see \cite{CHU}.

Our proposal however is more subtle. 

For the sake of definiteness, we will consider type IA theory. 
This can be obtained in two ways: as
T-dual of type IB , or as an orientifold of type IIA 
theory. In the second way, it
is from the very beginning a 10d theory if the initial IIA is; in the first,
it is of course compactified in at least one direction, and one can make
contact with the other approach by sending this radius to infinity. Either
way, we obtain a 10D theory (and no quantization on $B$). 
First we start with a brief review of some issues in  IA
theory.
The symmetry of IIA theory by which one orientifolds is $P_9 \cdot \Omega$, 
where $P_9: x^9 \to - x^9$ 
is a spacetime reflection and $\Omega: \sigma \to \pi -\sigma$ 
is worldsheet parity \cite{GP}. So, the orientifold plane is an eight-plane 
located at $x^9 =0$.
Physics in the $x^9 >0$ region is locally 
the same as the IIA one. However, strings always have an image  on the other 
side, and as a consequence spacetime fields are reflected as 
\be
\Phi (x^1, \ldots, x^8, x^9)= \pm \Phi(x^1, \ldots, x^8, -x^9), \label{refl}
\ee
the sign is determined by the $\Omega$ parity. So, in particular, the RR charges
of image branes have a relative $\pm$ according to their dimensionality. To 
obtain the
gauge groups we are looking for, namely $SO$ and $Sp$ groups, we have to put branes
and their mirrors on the
orientifold plane (otherwise gauge symmetry would be broken to $[U(n/2)]^2$); 
so, as far as we are concerned only branes whose mirrors have the same RR charge
survive -- the others meet their antibranes and annihilate. 
The surviving ones are 0, 4, and 8-branes; the gauge group on them is $SO$, $Sp$
and $SO$ respectively. From the low energy effective theory point of view, 
as stated in the literature, \cite{GSW},
the orientifold projection corresponds to charge conjugation operator.

Having specified that we have branes stuck on the orientifold plane, let us
now analyze more in detail the consequence of a (special) background $B$ field. 
Since we 
are really interested in its components parallel to the orientifold,  we set
$B_{\mu 9}=0, \mu=1\ldots 8$. As for the remaining components, 
bearing in mind that the $B$ field is
odd under worldsheet parity \cite{TASI}  from (\ref{refl}) we learn that 
$B_{\mu\nu} (x^1, \ldots, x^8, x^9) =
- B_{\mu\nu} (x^1, \ldots, x^8, - x^9)$. So,  we will consider a configuration 
$B_{\mu\nu} = b_{\mu\nu} f(x^9)$, 
where $f$ is odd in $x^9$. It is certainly true that
the $B$ field is zero on the orientifold, so this would seem hopeless; but 
strings which end on the branes can stretch also outside, and the usual
statement that their interaction with $B$ is a boundary term is, in general, 
true only when $B$ is constant. The interaction term equals ($\Sigma$ is the
worldsheet of the open string)
\bea 
\int_\Sigma B_{\mu\nu} d x^\mu \wedge d x^\nu =&& 
\int_\Sigma d \left( B_{\mu\nu} x^{\mu}d x^{\nu} \right) - 
\int_\Sigma dB_{\mu\nu}\wedge x^\mu d x^\nu \nonumber\\
&&=\int_{\partial\Sigma} B_{\mu\nu} x^\mu d x^\nu -
\int_\Sigma \partial_\rho B_{\mu\nu} d x^\rho \wedge x^\mu d x^\nu.
\eea
In the usual $B=$ constant case, 
the first term of the final expression is the boundary
term which is responsible for noncommutativity, while the second vanishes.
In our case, the situation is different: the first term is now zero, due to the
vanishing of the $B$ field on the orientifold plane (branes are on the
orientifold, so $\partial \Sigma \subset O8$), but the second is 
\be
\int_\Sigma \partial_9 f(x^9) b_{\mu\nu} x^\mu d x^\nu \wedge d x^9.
\ee
We choose $f$ to be the step function $\eps(x^9)$, which is in fact the
easiest field configuration one can think of in this case. The factor 
$\partial_9 f(x^9)$ becomes a $\delta (x^9)$, and this makes the integral to 
reduce of dimensionality and concentrate on the orientifold $\{ x^9 =0\}$:
\be
\int_{\Sigma\cap O8} b_{\mu\nu} x^\mu d x^\nu.
\ee
Now, as $\Sigma\cap O8 \supset \partial\Sigma$, this provides a boundary term
which has exactly the form of the one which usually accounts for
noncommutativity. 

This situation would seem at first 
to be strangely discontinuous with respect to what happens as
soon as one separates branes; in that case, one would be tempted intuitively
to consider boundary conditions which vary. For instance, for a string going
from one brane to its image, one would write boundary conditions 
$(g_{\mu\nu}\partial_n + \epsilon(y)b_{\mu\nu}\partial_y ) x^\nu =0$, where the
worldsheet is to be thought of as the upper half plane, $y$ is a
coordinate on the real axis, and $\partial_n$ denotes normal derivative. 
This is not right:
one has to remember that both $P_9$ and $\Omega$ have been applied. In detail:
the $\Omega P_9$ symmetry is expected to leave the closed string background 
$B_{\mu\nu} = 
\eps( x^9) b_{\mu\nu}$ 
invariant; to do so, since $x^9$ is reversed, the parameter
$b$ should be reversed as well. The operation $z \to -\bar z, b \to -b$ is
compatible with the usual boundary conditions  $(g_{\mu\nu}\partial_n + 
b_{\mu\nu}\partial_t ) x^\nu =0$, and not with the naive ones. 

One can see that all our arguments can also be extended to the case with
lower dimensional orientifold planes. On the contrary they cannot be extended
to the type IB case. However we do not exclude that \nc gauge theories may
arise in this case too.

\subsection{Supersymmetry argument.}

In this subsection we give another argument, based on supersymmetry, for
the stability of a system of 
$Dp$-branes on top of an orientifold plane ($Op$-plane with $p\leq 8$) in the 
presence of a step function-like $B$ field. To illustrate our reasoning we first 
consider the zero $B$ field case. 

Suppose that $Q_L$ and $Q_R$ represent the 32 supercharges of type II 
theories,
i.e. $Q_L$ and $Q_R$ have the same (or opposite) ten dimensional chiralities
for IIB (or IIA) theory. Introducing a $Dp$-brane, half of these 
supersymmetries are preserved. The corresponding conserved generators
are given by the following linear combination of $Q_L$ and $Q_R$, \cite{TASI}
\be\label{super}
Q=\epsilon_L Q_L +\epsilon_R Q_R,
\ee
where
\be\label{eps}
\Gamma^{01...p}\epsilon_L=\epsilon_R.
\ee
Now, let us consider an $Op$-plane parallel to the $Dp$-brane. To study the
stability of this system one way is to check whether the supercharges 
(\ref{super}),
or a portion of them, are preserved under the orientifold projection.
The $Op$-plane we are interested in is characterized by invariance under
$\Omega P_T$ operator, where 
$P_T$ is the parity in {\sl all} the directions transverse to the $Op$-plane.
Under $\Omega$, $Q_L$ and $Q_R$, which correspond to supersymmetries of
closed string left and right movers are interchanged. 
On the other hand under $P_T$, equation (\ref{eps}) is reversed, namely
$\Gamma^{01...p}\epsilon_R=\epsilon_L$. So, altogether 
$$
(\Omega\,P_T)\ Q=Q.
$$
In other words the presence of the parallel $Op$-plane does not break supersymmetry
any further and exactly the same supersymmetry as for a
$Dp$-brane is preserved. Hence the whole system is stable.
One can extend the above argument to an $Op$-plane parallel to a 
$D(p-4)$-brane.
Again this system is stable (it preserves some supersymmetry) however in this 
case 8 supercharges survive.

For the cases with non-zero $B$ field, we follow a similar discussion. 
For definiteness let us consider a rank one $B$ field, which is non-zero 
along $p-1$-th and
$p$-th directions; generalization to other cases is straightforward.  
The portion (half) of supersymmetry, which is preserved by the $Dp$--brane,
is given by \cite{SUSY} 
\be\label{superB}
\Gamma^{01...p}({1\over \sqrt{1+b^2}}-{b\over \sqrt{1+b^2}}\Gamma^{p-1,p})
\epsilon_L=\epsilon_R.
\ee

Now, we introduce the $Op$-plane, which again acts by $\Omega\ P_T$ projection.
Noting that
\begin{description}
\item [i.]$\Gamma^{01...p}$ and $(1-\Gamma^{p-1,p}b)$ are commuting, 
\item [ii.]$(1-\Gamma^{p-1,p}b)(1+\Gamma^{p-1,p}b)=(1+b^2){\bf 1}$ and
\item [iii.]under $\Omega\,P_T$,  $b$ is reflected to $-b$,
\end{description}
$\Omega\ P_T$, acting on (\ref{superB}), will lead to the same equation with
$\epsilon_L$ and $\epsilon_R$ interchanged. Therefore, there are 16 supercharges 
invariant under the $\Omega\ P_T$ transformation. Hence, our system formed 
by parallel $Dp$-branes in the 
background $B$ field introduced earlier, in the presence of a parallel 
$Op$-plane, is stable.
 
The above argument can also be understood in a  more intuitive way. 
A $Dp$-brane in a 
$b$ field (say $b_{p-1,p}$) background can be treated as a bound state of a $Dp$-  
and $D(p-2)$- branes, with $(p-2)$-branes having their worldvolume along the
$01...p-2$ directions, while $b_{p-1,p}$ represents their distribution density 
in the $p-1$ and $p$
directions \cite{{more}}. If we put this system in front of an $Op$-plane, as
above, $(p-2)$-branes are reflected to anti-$(p-2)$-branes; however since
at the same time we also change the $b$ to $-b$, the image of a $(p,p-2)$ bound 
state remains the same object, which is known to preserve 16 supercharges.   
 
Another argument (from which in fact (\ref{superB}) could follow) 
that supersymmetry is preserved can be obtained via a T duality 
transformation in a direction parallel to the branes. Suppose for 
example, in the $O8$ case, 
that our $p$--branes extend in the $3$ and $4$ directions,
and that there is a non-vanishing $b_{34}\equiv b$ field. Let us perform 
a T-duality in the direction $3$; by considering the boundary 
conditions before and after the transformation, it is easy to see that 
the branes get transformed to lower
dimensional tilted branes, which extend in the direction 
$x^4 - b x^3$. It is less trivial to understand what happens to the 
orientifold plane: to see what happens, we have to consider the 
transformation $\Omega P_9$ and interpret it in terms of the dual 
coordinates $x_D$:
$$
\begin{CD}
 x^\mu @>T_3>> x_D^\mu \\
 @V\Omega P_9VV   @VV\Omega P_\natural V\\
 \Omega P_9 x^\mu @>T_3>> (\Omega P_9x^\mu)_D.
\end{CD}
$$
The map $\Omega P_\natural$ is what we have to find, and is 
what defines the orientifold on the dual side. Let us consider the 
action on $x^3$; in this case $x^3_D =x^3(z) - x^3(\bar z)$. The map 
$\Omega P_9$ acts on the original expansion by sending 
$z\to -\bar z, b \to -b$ (and of course $x^9\to
-x^9$), and the $\Omega P_\natural$ map, 
on $x^3_D$, does the same but with an extra 
overall minus sign. This is not, however, the map $\Omega P_9 P_3$. 
On this side of the duality, $\Omega P_9 P_3$ (contrary to 
 $\Omega P_\natural$)
would indeed not touch $b$, since it is a number not a field as
seen from $\Omega P_9 P_3$ : it is in fact the angular coefficient 
by which the brane is tilted. To understand what the map really is, 
we have to act on the mixed coordinates $x^3 +b x^4, x^4 -b x^3$. These
have an expansion which contains only pure combinations 
$(z^{-n} - \bar z^{-n}), (z^{-n} + \bar z^{-n})$ respectively, and so 
in terms of the latter it is easily seen that $\Omega P_\natural$ is 
$\Omega P_9P_{x^3 + b x^4}$. This means
that the orientifold plane is tilted along the $x^4 - bx^3$ coordinate,
and so is parallel to the brane. The mirror brane is now necessarily 
also parallel to them, and the whole system preserves supersymmetry.

\subsection{Correlation functions.}

To show that actually the field theories we described above arise as low
energy effective actions on branes, we will now follow similar steps to the
usual $U(n)$ case. 
 
Low energy effective actions are found by computing scattering of strings
corresponding to the various effective fields. In our case, 
incoming states
have to be accompanied by orientifold projectors $(1+ \Omega P_T)/2$.  
This means that any vertex  $V_{\ket{s}}$ has to be changed in the combination
$ 1/2 (V_{\ket{s}} +V_{\Omega P_T \ket{s}})$, 
where the second term is the vertex
that creates the $\Omega P_T \ket{s}$ state. Let us specialize to the gauge
bosons, in which case the vertex, in the $-1$ picture, is $V(z) = 
\xi^{ij}\cdot (\psi +\bar \psi)e^{ikx}$, where we define
$\xi^{ij\,\mu} \equiv \xi^\mu \lambda^{ij}$. In the $B=0$ case,
since $\Omega P_T
(\xi\cdot b_{-1/2}\ket{0,k})=  - \xi^t\cdot b_{-1/2}\ket{0,k}$,  
correlation functions become
\be
\langle \
1/2 \left( V - V^t \right)(y_1) \ldots
1/2 \left( V - V^t \right)(y_n) \ \rangle,
\label{cofinal}
\ee
and give the usual result: amplitudes are obtained by substituting in 
the usual ones $\xi$ with $\xi - \xi^t$, i.e. keeping only the 
antisymmetric part of $\xi$.

In the present case, the analogy keeps working: now correlation 
functions after orientifolding are obtained from the ones before by 
the rule 
$\xi \to \xi -
\xi^{rt}$. Thus, for instance, for the gauge three point function  
the result is proportional to 
\bea
\Tr\left\{
(\xi_1- \xi^{rt}_1)\cdot p_2 \, (\xi_2- \xi^{rt}_2)\cdot
(\xi_3- \xi^{rt}_3)+
(\xi_2- \xi^{rt}_2)\cdot p_3 \, (\xi_3- \xi^{rt}_3)\cdot
(\xi_1- \xi^{rt}_1)+ \right.&&\nonumber \\
\left.(\xi_3- \xi^{rt}_3)\cdot p_1 \, (\xi_1- \xi^{rt}_1)\cdot
(\xi_2- \xi^{rt}_2)
\right\} e^{- \frac i2 p^1_\mu \theta^{\mu\nu}
p^2_\nu}+(1\leftrightarrow 2);&&
\eea
inner products are understood with respect to the open string
metric.
This is the same amplitude one finds starting from a \nc 
gauge theory, but with the additional constraint $\xi = - \xi^{rt}$; 
thus it coincides with the field theory we have suggested. 

As in theories arising from a non-orientifold case, there are two 
descriptions of the system that are equivalent at least in a 
perturbative sense, one is \nc and the other
commutative, \cite{SW}; the commutative one in the present case
is an ordinary $SO(n)$ gauge theory. So it is reasonable that, 
as we said, our theory is the image of a commutative $SO(n)$ gauge 
theory under the Seiberg-Witten map.
 
\section{Discussions and Remarks}

In this paper we have tackled the problem of \nc gauge theories, 
for groups other than $U(n)$. We have argued that to obtain the
\nc extension of a gauge theory, in general,
the usual interpretation of gauge symmetries as local internal symmetries 
should be modified. More precisely one must focus on 
an ``$NC$ Lie-algebra of gauge transformations", rather than on the
corresponding algebra of space-time independent transformations. 
Elaborating on the $NCU(n)$ group of gauge transformations, we have showed 
that actually one can extract some $NC$-subgroups of it. 
In particular we have discussed $NCSO(n)$
and $NCSp(n)$ gauge theories. As it is clear from our  
construction, in the commutative limit,
$\theta\to 0$, we recover the usual $SO$ and $Sp$ theories. 
Physically our method is based on the charge conjugation
operation in gauge theories. Noting the fact that the $SO(n)$ 
subgroup of the commutative $U(n)$ theory can be
constructed by simply choosing the gauge field to be in the charge 
conjugation invariant subgroup, it follows that one must 
restrict the gauge transformations to the same subgroup. The same idea
also works for the \nc case. But, of course, in the $NCU(n)$ case one 
has to consider the proper charge conjugation operator \cite{CPT}. 
Therefore, 
as we see, for the
particular case of $NCSO(2)$, this theory is not equivalent to $NCU(1)$,
although in the commutative case they are the same theory. The main 
difference between these two may be that the first, $NCSO(2)$, is 
invariant under the charge conjugation, but the other is not. So, 
in this way it seems more reasonable to consider the 
($NCSO(2)$ + fermions) as the proper \nc version of QED. From this 
example we also learn that given a commutative gauge theory, its \nc
extension  is not unique. Another special case is $NCSp(1)$. This 
theory can be treated as the \nc version for an $SU(2)$ gauge theory, 
and since there is no consistent way of finding \nc deformed $SU$ 
theories in general, $NCSp(1)$ seems to be very interesting.

{}From the string theory side of our construction,
since the orientifold projection corresponds to the charge conjugation 
operator in the low energy 
effective theory, we have guessed that a string theory environment that
might give rise to effective $NCSO(n)$ and $NCSp(n)$ theories should
contain an orientifold plane in a B-field background. Then,
calculating the corresponding scattering amplitudes,  we have
checked that this is indeed the case.
This provides a further support to our definition of  
$NCSO(n)$ and $NCSp(n)$ gauge theories. 

Naturally there are lots of interesting issues related to these 
theories, e.g. studying renormalizability,
which we do not deal with here but postpone to future research.   
However, we would like to end this paper with some remarks.
The first concerns chiral anomalies. In \cite{AS,gm,bst} the problem 
of anomaly cancellation in \nc $U(n)$ Yang--Mills theories was analyzed.
It was shown that anomaly cancellation can occur only by matching anomaly
coefficients from opposite chiralities. The main reason (although not the only one) is that the $U(1)$
factor does not decouple as in ordinary theories, therefore we cannot
define a \nc $SU(n)$ theory (as discussed in the introduction). This
fact motivated in part the research reported in this paper. One question
we would like to answer is: do there
exist \nc (non $U(n)$) gauge theories in which a more subtle 
cancellation mechanism for anomalies exists, with, in particular, anomaly--free
representations for chiral fermions? In this paper we have introduced new 
\nc gauge theories
with more general gauge groups. However, as far as anomaly cancellation
is concerned, the situation is no better, for example, in orthogonal theories 
than in $U(n)$ theories. The reason is that connections and gauge 
transformations have a nontrivial symmetric part. Therefore we are led 
back to the same conclusion as for \nc $U(n)$ theories. It would seem
that \nc Yang--Mills theories are definitely {\it more anomalous} than 
ordinary theories.

A second remark concerns the implications of $SO(n)$ being the Lorentz 
group in commutative $n$--dimensional spacetimes. Recently, its \nc
extension has been considered by gauging the $NCU(1,D-1)$ instead of 
the corresponding $SO$ symmetry \cite{Ali}. As a result, a
complexified gravity theory was found. Using our definition
of \nc $SO(n)$ gauge theory one can take, in regards to this problem
a different attitude. One can address the formulation of gravity 
theories on the Moyal plane, by using the $NCSO(1,D-1)$ gauge group
of transformations. This may lead to a more reasonable gravity theory, 
since it must correspond to the usual gravity theory when $\theta\to
0$. However, we expect that again in this case we will deal with some 
complexified gravity.

\vskip 2cm

{\bf Acknowledgements}

We are grateful to M. Bianchi, C.S.Chu, E.Gava and T. Krajewski
for useful discussions. M.M. Sh-J. would like to thank
N. Nekrasov for helpful remarks and K. Narain and A. Kashani-Poor for 
discussions. The research of L.B. , M.S. and A.T.   was partially 
supported by EC TMR Programme, grant FMRX-CT96-0012, and by the Italian 
MURST for the program ``Fisica Teorica delle Interazioni Fondamentali''.
The work of M.M. Sh-J. was partly supported by the EC contract 
no. ERBFMRX-CT 96-0090.

\end{document}